\documentclass[twocolumn,showpacs,preprintnumbers,amsmath,amssymb]{revtex4}
\usepackage{dcolumn}
\usepackage{bm}
\usepackage{graphicx}
\usepackage{amsmath}
\begin{document}

\title{Discrete chaotic states of a Bose-Einstein condensate}
\author{Wenhua Hai$^*$, \ \ Shiguang Rong, \ \ Qianquan Zhu}
\affiliation{Key Laboratory of Low-dimensional Quantum Structures
and Quantum Control of Ministry of Education, and \\ Department of Physics,
Hunan Normal University, Changsha 410081, China} 
\email{whhai2005@yahoo.com.cn}

\begin{abstract}

We find the different spatial chaos in a one-dimensional attractive
Bose-Einstein condensate interacting with a Gaussian-like laser
barrier and perturbed by a weak optical lattice. For the low laser
barrier the chaotic regions of parameters are demonstrated and the
chaotic and regular states are illustrated numerically. In the high
barrier case, the bounded perturbed solutions which describe a set
of discrete chaotic states are constructed for the discrete barrier
heights and magic numbers of condensed atoms. The chaotic density
profiles are exhibited numerically for the lowest quantum number,
and the analytically bounded but numerically unbounded Gaussian-like
configurations are confirmed. It is shown that the chaotic wave
packets can be controlled experimentally by adjusting the laser
barrier potential.

\end{abstract}


\maketitle

\textbf{}

\section{Introduction}

It is well-known that chaos in a nonlinear system not only may play
a destructive role, but also has many practical and dramatic
applications \cite{Kim}. Chaos has been thoroughly studied during
the last century in many different fields of physics. Very recently
it has been recognized that existence of chaos is also possible in
the Bose-Einstein condensates (BECs) described by the
Gross-Pitaevskii equation (GPE) picture \cite{Thommen} and in the
discretized systems describing trapped BECs within the space-mode
approximation \cite{Buonsante}. The temporal chaos was revealed in
the time evolutions of BECs trapped in a double-well potential
\cite{Abdullaev}. The spatial chaos with spatially disordered
configurations was investigated for the stationary states of the
BECs held in an optical lattice \cite{Eguiluz}. The spatiotemporal
chaos in BECs interacting with different potentials has also been
found \cite{chong1}.

The mean-field stationary states of a BEC are dominated by the
time-independent one-dimensional (1D) GPE \cite{Dalfovo, Leggett}.
For a BEC in ground state without current \cite{Dalfovo} the GPE is
a real equation and can be identical with the celebrated Duffing
equation \cite{Holmes, Parthasarathy} by using time instead of
spatial coordinate. Particularly, when such a GPE is perturbed by a
weak periodic potential, the Smale-horseshoe chaos may appear for a
certain parameter region of the extended dynamical system
\cite{Holmes, Parthasarathy}. The Melnikov chaos criterion gives the
chaotic parameter region in which the perturbation parameters are
allowed to vary their values continuously \cite{Melnikov, Liu,
Gukenheimer,Hai1}. The Gaussian-like barrier potentials can be
realized by a sharply focused laser beam in the experiment
\cite{Burger}, which have been applied to investigate the shock-wave
formation in BECs \cite{Simula}, the nonlinear resonant transport
\cite{Paul} and deterministic chaos \cite{Coullet} of BECs.
Recently, using external fields to control quantum states of BECs
has become an important physical motivation \cite{Creffild}.

When a BEC is created initially in a time-independent optical
lattice, the stationary states of the GPE are determined by the
boundary conditions and are adjusted by the system parameters. The
different boundary conditions may be established in a practical
experiment, which cannot be set accurately. In the non-chaotic
regimes, a small change of the boundary conditions and/or system
parameters brings the BEC state only a small correction which can be
neglected in a good approximation. In the chaotic regimes, however,
the stationary state depends on the conditions and parameters
sensitively. The sensitivity means that a small change of the
conditions and/or parameters may cause a great difference which is
not negligible. For example, the periodic configuration of BEC
density is changed to the aperiodic and irregular one. It is
important for the application purpose to predict the bounded states
and to manipulate the corresponding density distributions which
govern the beam profile of an atom laser extracted from the BEC
\cite{Kohl}. Therefore, investigating the spatial chaos and its
control is necessary and interesting for the considered BEC system.

The main aim of this paper is to present an analytical evidence of a
different type of spatial chaos which can be defined as the discrete
chaotic states, and to establish a method for controlling the
chaotic states. By the discrete states we mean a denumerable set of
bounded solutions in which any solution is one-to-one with a value
in a discrete set of the parameter values. If the discrete states
meet the Melnikov chaos criterion, we call them the discrete chaotic
states. By using a laser beam modeled by the tanh-squared-shaped
barrier potential \cite{Wang} which is known as the Rosen-Morse
potential \cite{Lamb}, we demonstrate the existence of spatial chaos
in the BEC held in a weak optical lattice. The chaotic regions of
parameters are exhibited and the regular and disordered
configurations of the BEC are illustrated. It is shown that the
width and site of the strong barrier potential confine the width and
site of the BEC wave-packet, and a denumerable set of the barrier
height values corresponds to the discrete chaotic states and magic
numbers of condensed atoms. Thus the possible chaotic states can be
controlled by adjusting the width, site and height of the laser
barrier experimentally.

\section{Chaotic and regular states for the low laser barrier}

For the considered BEC system with transverse wave function being in
ground state of a harmonic oscillator of frequency $\omega_r$, the
governing time-independent quasi-1D GPE reads
\begin{eqnarray}
- \frac{\hbar^2}{2m}\psi _{xx} + [V'(x)+ g'_{1d} |\psi|^2]\psi=\mu
\psi,
\end{eqnarray}
where $m$ is the atomic mass, $\mu$ is the chemical potential, and
$g'_{1d}= g_0 m \omega_r/(2\pi \hbar)= 2 \hbar\omega_r a_s$ denotes
the quasi-1D atom-atom interaction intensity with $a_s$ being the
$s$-wave scattering length. Hereafter, by $\psi_{xx}$ we mean the
second derivative of $\psi$ with respect to $x$. The external
potential $V'(x)=-V_0\tanh^2 [\beta (x-x_c)]+ V_1\sin^2 kx$ contains
the longitudinal barrier potential of strength $V_0>0$, width
$\beta^{-1}$ and center site $x_c$, and the perturbed lattice
potential with $V_1$ and $k$ being the intensity and wave vector.
The former as a Gaussian-like potential can be formed by a sharply
focused laser beam in the experiment \cite{Burger}, and the latter
is a laser standing wave. Taking $\beta^{-1}$ and $\beta$ as the
units of coordinate $x$ and density $|\psi|^2$, and normalizing the
potential strengths $V_0, V_1$ and chemical potential $\mu$ by using
$E_{\beta}=\hbar^2\beta^2/m$, Eq. (1) becomes the dimensionless
equation
\begin{eqnarray}
- \frac{1}{2}\psi _{xx} + [V(x)+ g_{1d} |\psi|^2]\psi=\mu\psi.
\end{eqnarray}
Here the interaction intensity is reduced to $g_{1d}= 2 \hbar
\omega_r a_s\beta/E_{\beta}=2a_s/(\beta a_r^2)$ with
$a_r=\sqrt{\hbar/(m\omega_r)}$ being the transverse harmonic
oscillator length, and the potential gets the form
\begin{eqnarray}
V(x)=-V_0\tanh^2 (x-x_c)+ V_1\sin^2 kx
\end{eqnarray}
with $k$ measured in $\beta$.

We are interested in the real solution of GPE (2), which makes the
GPE the perimetrically perturbed Duffing equation
\cite{Parthasarathy} in the spatial evolution and for the weak
potential. It is well known that existence of the periodic
perturbation is necessary for the appearance of chaos in the Duffing
system \cite{Melnikov, Liu, Gukenheimer}. When negative interaction
and negative chemical potential are taken, in the absence of
external potential the system has the well-known homoclinic
(separatrix) solution \cite{Parthasarathy, Melnikov, Liu,
Gukenheimer}
\begin{eqnarray}
\psi_0&=&\sqrt{\frac{2\mu}{g_{1d}}}\ \textrm{sech}[\sqrt{-2\mu}(x-c_0)],\nonumber \\
C_0&=&\frac{1}{\sqrt{-2\mu}}\Big\{x_0-\textrm{Arc
sech}\Big[\sqrt{\frac{g_{1d}}{2\mu}}\psi_0(x_0)\Big]\Big\},
\end{eqnarray}
where $c_0$ is an arbitrary constant adjusted by the boundary
conditions at the boundary $x=x_0$. For the BEC system governed by
Eq. (2) the constant $c_0$ cannot be determined experimentally,
because of the undetectable $\psi_0(x_0)$. The presence of the weak
external potential leads to the Melnokov function
\cite{Parthasarathy, Melnikov, Liu, Gukenheimer}
\begin{eqnarray}
& & M(c_0)=\int_{-\infty}^{\infty}
2\psi_{0x}(x)V(x)\psi_0(x)dx\nonumber
\\
&=&\frac{4\mu\sqrt{-2\mu}}{g_{1d}}\Big[F- \frac{k^2\pi
V_1\sin(2kc_0)}{2\mu\sinh(k\pi/\sqrt{-2\mu})}\Big]
\end{eqnarray}
for $0<V_0\ll 1$ and $|V_1|\ll 1$, where $\psi_{0x}$ denotes the
first derivative of $\psi_0$ with respect to $x$, constant $F$ from
the barrier potential reads
\begin{eqnarray}
F=\frac{4V_0 e^{2\Delta}}{|\mu|(e^{2\Delta}-1)^4}[(3+2\Delta
+8\Delta e^{2\Delta}+(2\Delta-3) e^{4\Delta})]
\end{eqnarray}
with $\Delta=c_0-x_c$. The Melnokov function measures the distance
between the stable and unstable manifolds in the Poincar\'{e}
section of the equivalent phase space $(\psi,\ \psi_x)$. For some
$c_0$ values if the Melnikov function has a simple zero, the locally
stable and unstable manifolds intersect transversally such that the
Smale-horseshoe chaos exists in the Poincar\'{e} map
\cite{Parthasarathy, Melnikov, Liu, Gukenheimer}. The possibility of
$M(c_0)=0$ results in the chaotic region of parameter space
\begin{eqnarray}
|V_1|\ge 2\frac{|F\mu|}{\pi
k^2}\sinh\Big(\frac{k\pi}{\sqrt{-2\mu}}\Big).
\end{eqnarray}
When parameters are taken in the chaotic region, the Melnikov
function has zero points and the stable and unstable manifolds in
the Poincar\'{e} section may intersect that leads to the
Smale-horseshoe chaos. It is possible that the regular orbits exist
for both the chaotic and non-chaotic regions. The chaotic and
regular orbits in the chaotic region depend on the different
boundary conditions respectively.

\begin{figure}
 \center
\includegraphics[width=1.6in]{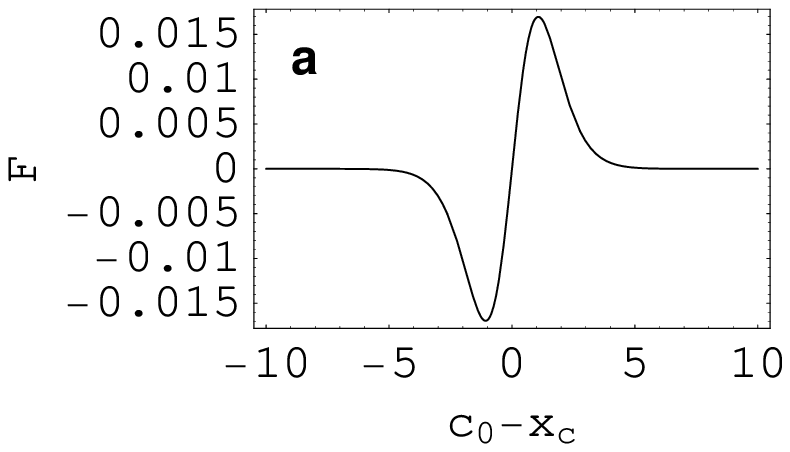}
\includegraphics[width=1.6in]{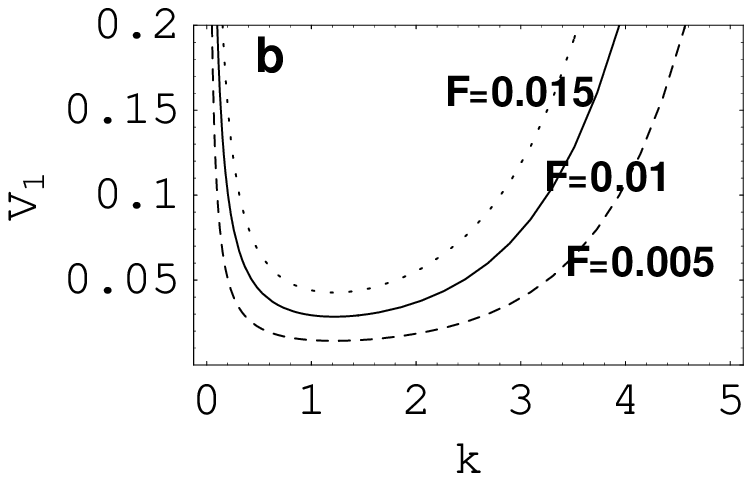}
\caption{(a) The constant $F$ as a function of $c_0-x_c$ for
parameters $\mu=-2$ and $V_0=0.2$. (b) The boundaries of the chaotic
regions for $\mu=-2$, and $F=0.005$ (dashed curve), $F=0.01$ (solid
curve) and $F=0.015$ (doted curve).}\label{fig1}
\end{figure}
\begin{figure}
 \center
\includegraphics[width=2.0in]{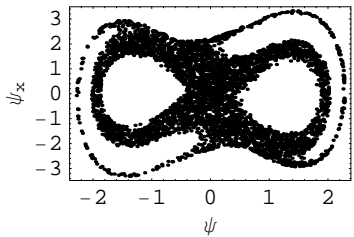}
\caption{The Poincar\'{e} section on the equivalent phase space
$(\psi,\ \psi_x)$ from Eq. (2) for the given parameters and boundary
condition. }\label{fig2}
\end{figure}
As can be seen from Eq. (7), for any negative chemical potential
$\mu<0$ and any barrier potential strength in the region $0<V_0\ll
1$, the chaotic region depends on constant $F$ in the plane of
parameters $V_1$ versus $k$. The $F$ is determined by the parameters
$V_0,\ \mu$ and $c_0-x_c$ with the potential strength $V_0$ and site
$x_c$ being adjustable. In Fig. 1a we show $F$ as a function of
$c_0-x_c$ for $\mu=-2$ and $V_0=0.2$ by using the MATHEMATICA code.
From this figure it can be observed that $|F|$ has a maximum
$|F|=0.015$ and a minimum $|F|=0$. The former corresponds to the
minimal chaotic region of Eq. (7), and the latter is associated with
the maximal chaotic region $|V_1|>0$. Taking $\mu=-2$ and $F=0.005,\
0.01,\ 0.015$ associated with three different $c_0$ values
respectively, from Eq. (7) we plot the boundary curves of the
chaotic regions as the dashed curve, solid curve and doted curve of
Fig. 1b. The corresponding chaotic regions are above these curves
respectively. The minimal chaotic region above the curve of
$|F|=0.015$ is certainly chaotic region for arbitrary $c_0$ value.
But the other chaotic regions are related to the corresponding
boundary conditions, through the constant $F(c_0)$.

\begin{figure*}
 \center
\includegraphics[width=1.7in]{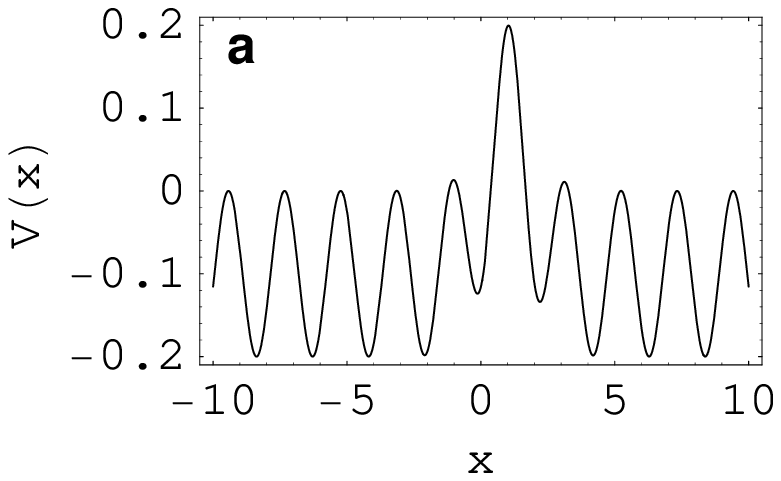}
\includegraphics[width=1.7in]{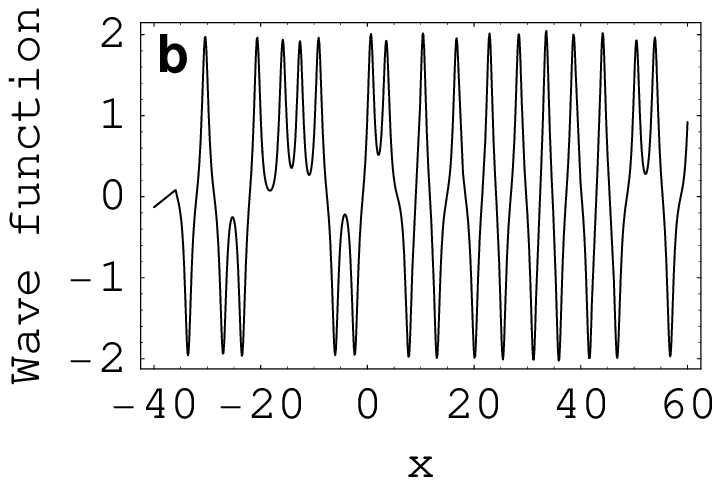}
\includegraphics[width=1.7in]{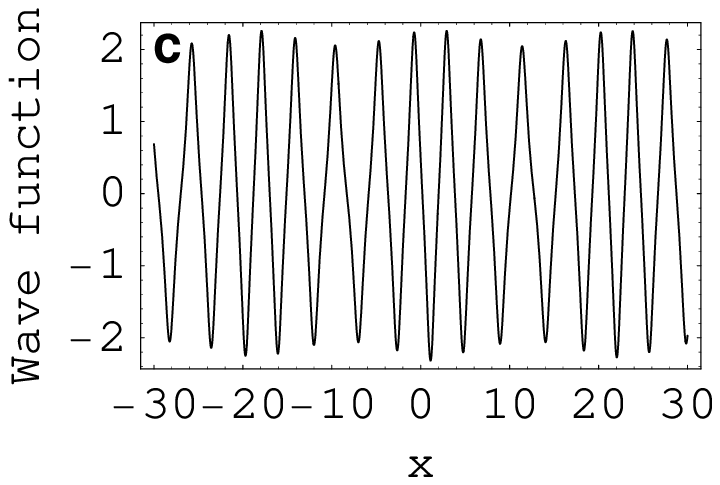}
\caption{(a) The potential function of Eq. (3) and (b) the aperiodic
chaotic state of Eq. (2) for the same parameters and boundary
condition with Fig. 2. (c) When the value of $\psi(10000)$ is
changed from $0.00001$ to $0$, we get the periodic wave
function.}\label{fig3}
\end{figure*}

\begin{figure*}
\center
\includegraphics[width=1.7in]{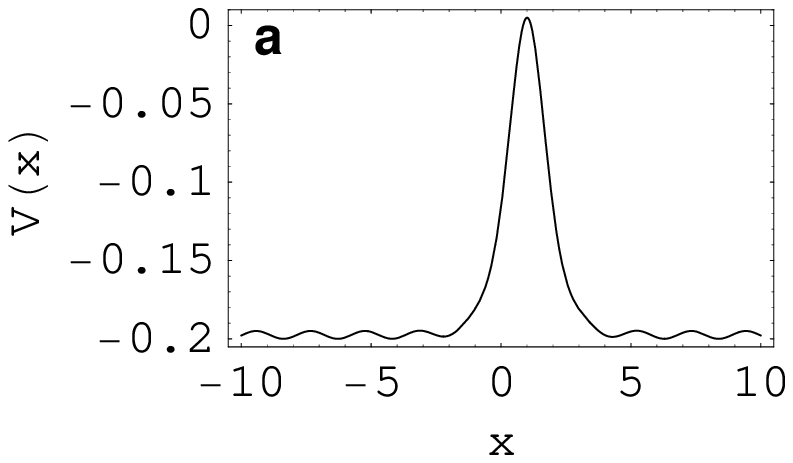}
\includegraphics[width=1.7in]{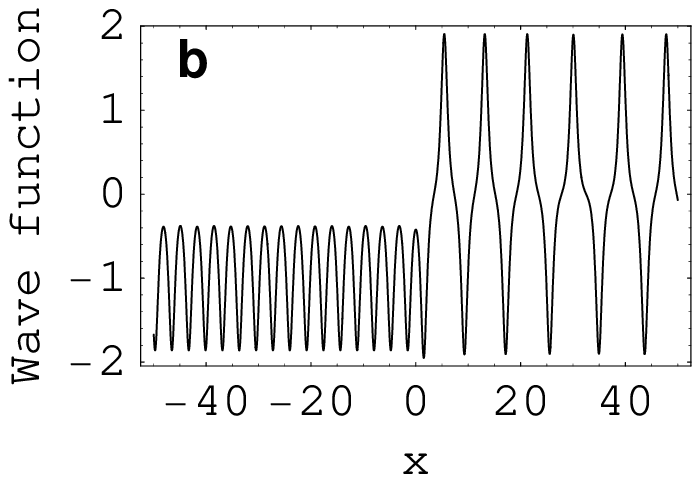}
\includegraphics[width=1.7in]{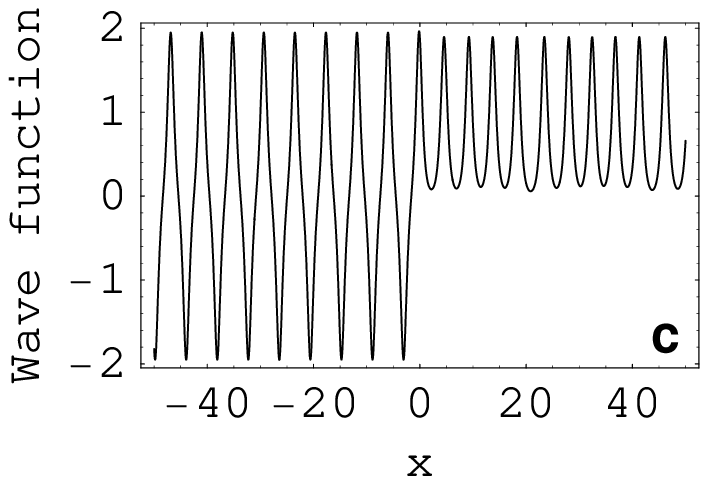}
\caption{The correspondences of Fig. 3 after the change of parameter
$V_1$ from $0.2$ to $0.005$.}\label{fig4}
\end{figure*}
A useful way of analyzing chaotic motion is to look at what is
called the Poincar\'{e} section, which is a discrete set of the
phase space points at every period of the periodic potential, i.e.
at $x = 2\pi/k, 4\pi/k, 6\pi/k, \cdots$. Taking the parameters
$\mu=-2,\ g_{1d}=-1,\ V_0=0.2,\ V_1=0.2,\ k=1.5,\ x_c=1$ and the
approximation $[\psi(x_0),\psi_x(x_0)]=[\psi(10000),
\psi_x(10000)]=(0.00001,0.00001)$ to the experimentally possible
boundary condition $[\psi(\infty), \psi_x(\infty)]=(0,0)$, from Eq.
(2) we numerically plot the Poincar\'{e} section on the equivalent
phase space $(\psi,\ \psi_x)$ and find the chaotic trajectory as in
Fig. 2. Here the lattice strength $V_1$ and wave vector $k$ are
evaluated in the minimal chaotic region of Fig. 1b. For the same
parameters of Fig. 2 from Eqs. (2) and (3) the potential and chaotic
state functions are plotted as in Figs. 3a and 3b respectively. From
Fig. 3a we can see the profile of the combined potential between the
barrier potential and periodic lattice. In Fig. 3b we exhibit the
aperiodicity and irregularity of the chaotic macroscopic wave
function corresponding to Fig. 2 numerically. In order to confirm
the sensitive dependence of chaotic system on the boundary
conditions, we change only the boundary condition as $[\psi(10000),
\psi_x(10000)]=(0,\ 0.00001)$ to plot the wave function. This small
change leads the irregular curve in Fig. 3b to the periodic one in
Fig. 3c. When the lattice strength is decreased to $V_1=0.005$ and
the other parameters are kept, from Fig. 1b we observe that the
parameter value is outside the given chaotic region. After changing
$V_1$ from $0.2$ to $0.005$, Figs. 3a, 3b and 3c are changed to
Figs. 4a, 4b and 4c respectively. Figure 4a displays the weak
periodic potential compared to the laser barrier. In Fig. 4b and 4c
we illustrate that in the considered parameter region the wave
functions are periodic for the given boundary conditions. It is
interesting noting that the regular wave functions in Figs. 4b have
two different periods and two different amplitudes in both sides of
the laser barrier. This means that the atomic number
$\int_{\Sigma}|\psi(x)|^2dx$ is different for the integration region
$\Sigma$ of different side. The periodicity is varied with the
change of the boundary conditions from $[\psi(10000),
\psi_x(10000)]=(0.00001,0.00001)$ of Fig. 4b to $[\psi(10000),
\psi_x(10000)]=(0,\ 0.00001)$ of Fig. 4c. Differing from Fig. 4b, in
Fig. 4c the period of wave function in both sides of barrier is the
sameness and the smaller one of amplitudes is enlarged compared to
that of Fig. 4b. The results display the different profiles of
macroscopic quantum states and reveal that the existence of chaos
means the sensitive dependence of the BEC system on the boundary
conditions and parameters.

\section{Discrete chaotic states for the high laser barrier}

The chaotic region of Eq. (7) is based on the perturbation theory
\cite{Melnikov, Liu} so that it is valid only for very small
potential strengths $V_0$ and $V_1$. When the strength $V_0$ of the
barrier potential is continuously increased to strong enough, e.g.
$V_0 >1$, it can no longer be treated as a part of perturbations. In
this case we require to reconsider the perturbation problem of the
stationary states. Applying the well-known Rayleigh-Schr\"{o}dinger
expansions \cite{Zeng}
\begin{eqnarray}
\psi=\psi_0+\psi_1, \ \mu=\mu_0+\mu_1 \ \ \ \textrm{for} \ \
|\psi_1|,\ |\mu_1|,\ |V_1|\ll 1
\end{eqnarray}
to Eq. (2) of real $\psi$, we have the leading order and the first
order equations as
\begin{eqnarray}
- \frac{1}{2}\psi _{0xx} &-& [V_0\tanh^2 (x-x_c)- g_{1d}
\psi_0^2]\psi_0=\mu_0\psi_0,
\\
- \frac{1}{2}\psi _{1xx} &-& [V_0\tanh^2 (x-x_c)- 3g_{1d}
\psi_0^2+\mu_0]\psi_1 \nonumber \\ =(\mu_1&-&V_1\sin^2kx)\psi_0(x).
\end{eqnarray}
Noticing that Eq. (9) has many special solutions for the fixed
values of $V_0,\ g_{1d},\ x_c$ and different $\mu_0$ values. Only
the homoclinic solution is related to the Melnikov's chaos and the
other solutions are associated with the regular states of Eq. (2).
Here we are interested in the chaos and only consider the homoclinic
solution thereby. It can be easily proved that the homoclinic
solution of Eq. (9) has the form
\begin{eqnarray}
\psi_0=\sqrt{\frac{V_0+1}{-g_{1d}}} \textrm{sech}(x-x_c) \
\textrm{for} \ \mu_0=-V_0-\frac 1 2.
\end{eqnarray}
Differing from Eq. (4), Eq. (11) describes a wave packet whose
height and width are adjusted by the potential intensity $V_0$ and
width $\beta^{-1}$ implied in the unit of $x$. Substituting Eq. (11)
into Eq. (10) yields the non-homogeneous equation
\begin{eqnarray}
- \frac{1}{2}\psi _{1xx} &-& \Big[(2V_0+3)\textrm{sech}^2
(x-x_c)-\frac{1}{2}\Big]\psi_1 \nonumber
\\ =-f(x)&=&(\mu_1-V_1\sin^2kx)\psi_0(x).
\end{eqnarray}
The corresponding homogeneous equation for $f=0$ is a well-known
Schr\"{o}dinger one with trapping potential
$-(2V_0+3)\textrm{sech}^2 (x-x_c)$ and eigenenergy $E=-1/2$. Given
two linearly independent solutions of the homogeneous equation as
$\psi'_1$ and $\psi''_1=\psi'_1\int (\psi'_1)^{-2}dx$, the exact
general solution of non-homogeneous Eq. (12) can be written in the
integral form \cite{Hai2}
\begin{eqnarray}
\psi_1=2\psi''_1\int_A^x \psi'_1f(x)dx-2\psi'_1\int_B^x
\psi''_1f(x)dx,
\end{eqnarray}
where $A$ and $B$ are arbitrary constants determined by the boundary
and normalization conditions. This solution can be directly proved
by comparing the second derivative $\psi _{1xx}$ from Eq. (13) with
that in Eq. (12).

Boundedness of the perturbed correction $\psi_1$ is the physical
requirement, which depends on the bounded $\psi'_1$. In order to
seek such a $\psi'_1$, we set \cite{Zeng}
\begin{eqnarray}
\psi'_1&=&[\textrm{sech}(x-x_c)]^{2\lambda}u(z), \ z=-\sinh^2(x-x_c), \nonumber \\
\lambda&=&[\sqrt{8(2V_0+3)+1}-1]/4.
\end{eqnarray}
Inserting Eq. (14) into the homogeneous part of Eq. (12) with $f=0$
produces the hypergeometric equation
\begin{eqnarray}
z(1-z)u_{zz}+[0.5-(a+b+1)z]u_z-abu=0,
\end{eqnarray}
where $a=0.5-\lambda, \ b=-0.5-\lambda$. Its two linear independent
solutions with finite terms read \cite{Zeng}
\begin{eqnarray}
u_n^e&=&F(0.5-\lambda,-0.5-\lambda,0.5,z) \ \ \textrm{for} \ \
\lambda
=0.5+n;\nonumber\\
u_n^o&=&\sqrt{|z|}F(1-\lambda,-\lambda,1.5,z) \ \ \textrm{for} \ \
\lambda =1+n.
\end{eqnarray}
Here $F(a,b,c,z)$ is the hypergeometric function, $u_n^e$ and
$u_n^o$ with $n=0,1,2,\cdots$ denote even and odd functions of
$(x-x_c)$ respectively. Combining Eq. (16) with Eq. (14) we arrive
at the bounded solutions
\begin{eqnarray}
\psi^{'e}_{1n}=[\textrm{sech}(x-x_c)]^{1+2n}u_n^e,
V_0=[(3+4n)^2-25]/16,\nonumber \\
\psi^{'o}_{1n}=[\textrm{sech}(x-x_c)]^{2+2n}u_n^o,
V_0=[(5+4n)^2-25]/16\nonumber
\\ \ \ \ \ \textrm{for} \ \ \ \ V_0>0, \ \ n=1,2,\cdots.\ \ \ \ \ \ \ \ \ \
\ \ \ \ \ \
\end{eqnarray}
Noting that $\psi'_1$ of Eq. (17) tends to zero, and
$\psi''_1=\psi'_1\int (\psi'_1)^{-2}dx$ of Eq. (13) is infinity at
$x=\pm\infty$. Thus the second term of Eq. (13) is in the form of
zero multiplying infinity at $x=\pm\infty$, so we can use the
l'Hospital rule to calculate the limit and to prove the boundedness
of this term. For the first term of Eq. (13) we have to establish
the boundedness condition
\begin{eqnarray}
I_{\pm}=\lim_{x \to\pm\infty}\int_A^x \psi'_1
(V_1\sin^2kx-\mu_1)\psi_0dx=0.
\end{eqnarray}
The necessity of Eq. (18) is obvious for the boundedness of Eq.
(13), because of the unboundedness of $\psi''_1$. Under condition
(18) we can apply the l'H\"{o}pital rule to the both terms of Eq.
(13), obtaining \cite{Hai2} $\lim_{x \to\pm\infty}\psi_1=2\lim_{x
\to\pm\infty}f(x)=0$. This limit implies that Eq. (18) is also
sufficient and the obtained macroscopic wave function satisfies the
usual boundary condition $\psi(\pm \infty)=\psi_0(\pm
\infty)+\psi_1(\pm \infty)=0$. Noticing the correspondence between
$f(x)$ and $\varepsilon _k^{(1)}(x)$ in Eq. (9) of the second
article of Ref. \cite{Hai2}, the above proof of sufficiency is
clear.

The integration of the first term in Eq. (13) is insolvable and
cannot be expressed by finite elementary functions. Hence, in the
numerical computation based on Eq. (13), small deviation from the
exact value of the integration satisfying condition (18) is
avoidable. The small deviation will be amplified exponentially fast
by the unbounded function $\psi''_1(x)$ until infinity as $x\to\pm
\infty$ that exhibits the numerical instability. The analytical
insolvability and numerical instability can cause the unpredictable
chaotic behavior \cite{Hai1}. The difference $I_+-I_-$ of the
integration in Eq. (18) is similar to the Melnikov function of Eq.
(5), hence $I_+-I_-=0$ can be called the generalized Melnikov
criterion for chaos. In fact, given Eq. (18), the undetermined form
$\psi''_1(\pm\infty)\times I_{\pm}=\infty\times 0$ appears in Eq.
(13) as $|x|$ tending to infinity, which leads to a new feature as
the analytical boundedness but numerical unboundedness, namely the
evidenced incomputability and unpredictability of the chaotic
behavior \cite{Hai1}. Therefore, under the condition (18) the
solution $\psi(x)=\psi_0(x)+\psi_1(x)$ in terms of Eqs. (11) and
(13) is called the chaotic solution \cite{Hai1}. If the zero
boundary condition $[\psi(\pm \infty),\psi_x(\pm \infty)]=(0,0)$ is
required theoretically, the uniqueness theorem infers the chaotic
solution to be the unique one of the system. On the other hand, from
the formula of the energy functional \cite{Dalfovo, Leggett},
\begin{eqnarray}
H&=&\int \psi^+ \Big[-\frac 1 2 \nabla^2+V(\vec{r})+\frac
{1}{2}g_{1d}|\psi|^2\Big]\psi d^3x \nonumber \\ &=& \int \Big[\frac
1 2 |\nabla \psi|^2+ V(\vec{r})|\psi|^2+\frac
{1}{2}g_{1d}|\psi|^4\Big]d^3x\nonumber \\ &-&\frac 1 2
(\psi^+\nabla\psi) |_{-\infty}^{\infty},
\end{eqnarray}
we know that unlike the unbounded solution with $|\psi^+(\pm
\infty)|=\infty$, the analytically bounded solution with $\psi^+(\pm
\infty)=0$ is associated with the finite energy functional and may
be metastable thereby \cite{Dalfovo}. Although the chaotic solution
is not very stable, due to the sensitive dependence on the
parameters and boundary conditions, it may also be metastable
compared to the analytically unbounded solution. \emph{Particularly,
these bounded solutions are valid only for the discrete $V_0$ values
of Eq. (17). This means the corresponding analytically bounded
chaotic states to be discrete with the increase of the barrier
height.}

The above-mentioned results imply that when the barrier potential is
strong enough, its strength values must be discrete for the bounded
perturbed solutions. For the discrete $V_0=V_{0n}$ values the
leading number-density $\psi_0^2$ is proportional to $V_{0n}$ and
the leading chemical potential is given as $\mu_{0n}=-\frac 1 2
-V_{0n}$ by Eq. (11), the both are also discrete. The parameters
$V_1, k, x_c$ and $g_{1d}$ can vary their values continually in a
certain parameter regions. Given a set of values of $V_1, k, x_c$,
the first correction $\mu_1$ is determined by the boundedness
condition of Eq. (18). In Eq. (10) the discrete chemical potential
$\mu\approx \mu_{0n}+\mu_1$ is equivalent to the energy of a
Schr\"{o}dinger system. In quantum mechanics \cite{Zeng}, it is
known that the boundedness of wave function may lead the energy to
take discrete values. Mathematically, the relationship between the
discrete values of potential strength $V_0=V_{0n}$ and the exactly
bounded solutions of Eq. (12) agrees qualitatively with that of a 2D
Coulomb correlated system \cite{Taut}, where the Schr\"{o}dinger
equation is exactly solvable only for a denumerably infinite set of
values of magnetic strength (or the corresponding oscillator
frequency). Physically, we well know that for a 2D electron gas in a
semiconductor heterojunction the integral and fractional quantum
Hall plateaus are associated with the discrete set of values of
magnetic strength \cite{Prange}.

We now investigate the physical effect of the discrete laser
strength $V_0=V_{0n}$ on the considered BEC system. Applying Eq.
(11) to the normalization condition yields the number of condensed
atoms $N_n\approx \int |\psi_{0n}|^2dx= 2(1+V_{0n})/|g_{1d}|
=(1+V_{0n})\beta a_r^2/|a_s|$ for the metastable states given by Eq.
(8) with Eqs. (11) and (17), that results in the relation
\begin{eqnarray}
N_n|a_s|\approx (1+V_{0n})\beta a_r^2
\end{eqnarray}
with $V_{0n}$ given in Eq. (17). Here the special value $N_n$ can be
called the magic numbers of the macroscopic many-body system keeping
in the metastable states. Differing from the magic numbers of the
microscopic many-body system (e.g. atomic nucleus), $N_n$ denotes
some approximate values, because of the approximation $N\pm 1\approx
N$ in the mean-field theory of macroscopic many-body system
\cite{Dalfovo,Leggett}. For a harmonically confined BEC system, the
supercritical number $N_{cr}$ of condensed atoms obeys
\cite{Dalfovo} $N_{cr}|a_s|=0.575 a_{ho}$ with $a_{ho}$ being the 3D
harmonic oscillator length. The magic number $N_n$ may exceed the
supercritical number $N_{cr}$ by increasing the laser strength
$V_{0n}$ and/or decreasing the laser barrier width $\beta^{-1}$. The
approximate magic numbers of the considered many-body system
warrants experimental investigation.

Let us take the simplest even solution of $(x-x_c)$ with quantum
number $n=1$ as an example to show the feature of the chaotic
solutions. From Eqs. (17) and (16) such a solution is derived as
\begin{eqnarray}
\psi'_{1n}=\psi^{'e}_{11}=\textrm{sech}^3y(1-4\sinh^2y)
\end{eqnarray}
for $y=x-x_c$ and $V_{01}=3/2,\ \mu_{01}=-1/2-V_{01}=-2$. Obviously,
this solution tends to zero as $x\to\pm\infty$. The corresponding
unbounded solution reads
\begin{eqnarray}
& & \psi''_{1n}=\psi^{''e}_{11}=\psi^{'e}_{11}\int
(\psi^{'e}_{11})^{-2}dx \nonumber
\\ &=&\frac{1}{64}\textrm{sech}^3y(36y-24y\cosh 2y+28 \sinh 2y-\sinh 4y) \nonumber
\\
\end{eqnarray}
in which the term $\frac{1}{64}\textrm{sech}^3y\sinh 4y$ tends to
$\pm\infty$ and the other terms tend to zero as $x\to\pm\infty$.
Applying Eqs. (21) and (22) to Eq. (13), the exact general solution
of Eq. (12) becomes
\begin{eqnarray}
\psi^{e}_{11}=2\psi^{''e}_{11}\int_A^x
\psi^{'e}_{11}f(x)dx-2\psi^{'e}_{11}\int_B^x \psi^{''e}_{11}f(x)dx,
\end{eqnarray}
where $f(x)=-(\mu_1-V_1\sin^2kx)\psi_0(x)$ is equal to zero at
$x=\pm\infty$, because of $\psi_0(\pm\infty)=0$. Applying the
l'H\"{o}pital rule to Eq. (23), we easily verify its boundedness
\cite{Hai2}, through the limit $\lim_{x\to
\pm\infty}\psi^{e}_{11}(x)=0$ for the $\mu_1$ obeying Eq. (18)
accurately. Inserting $\psi^{'e}_{11}$ and Eq. (11) into Eq. (18),
from the generalized Melnikov chaos criterion $I_+-I_-=0$ one
derives
\begin{eqnarray}
\mu_1=0.5\pi V_1\cos(2kx_c)k(5k^2-1)\textrm{csch}(k\pi)
\end{eqnarray}
which can be adjusted by the laser site $x_c$ and has a maximum and
a minimum at $\cos(2kx_c)=\pm 1$ respectively.

In order to obtain the bounded numerical solution of Eq. (23), the
parameter $\mu_1$ must obey Eq. (24). However, in any numerical
computation, for a set of fixed parameters $V_1,\ k,\ x_c$ it is
impossible to take the value of $\mu_1$ accurately, because of the
irrational $\pi$ with infinite sequence of digits in Eq. (24). This
implies small deviation from the accurate boundedness condition (18)
and the small deviation will lead the numerical solution of Eq. (23)
to be exponentially amplified by the unbounded function
$\psi''^e_{11}$ until infinity as $x\to\pm \infty$. So the
analytically bounded chaotic solution (23) is numerically unbounded
and uncomputable for sufficiently large $|x|$ values \cite{Hai1}.
For a small $|x|$ value $\psi''^e_{11}$ is finite and Eq. (23) is
certainly bounded. At $x=\pm \infty$ the boundedness condition (18)
and l'H\"{o}pital rule lead Eq. (23) to zero analytically. The
unpredictability of chaotic solution (23) may occur only near the
spatial range $|y|=|x-x_c|\in (|y_s|,\ \infty)$, where $y_s=x_s-x_c$
can be estimated through the starting point of the numerical
incomputability after which the solution tends to infinity rapidly.
In such a spatial range, the chaotic region of atomic density may be
$|\psi (y)|^2 \in (2|\psi_0 (\pm \infty)\psi^{e}_{11}(\pm \infty)|,\
2|\psi_0 (y_s)\psi^{e}_{11}(y_s)|)=(0,\ 2|\psi_0
(y_s)\psi^{e}_{11}(y_s)|$ with width $\delta(y)$ tending to zero as
the increase of $|y|$ value. The maximal width reads
$\delta(y_s)\approx 2|\psi_0 (y_s)\psi^{e}_{11}(y_s)|$ which is in
order of perturbation $V_1$, since $|\psi(y)|^2$ equates
$|\psi_0(y)+\psi^{e}_{11}(y)|^2\approx |\psi_0(y)|^2+
2|\psi_0(y)\psi^{e}_{11}(y)|$ and $|\psi_0(y)|^2$ is predictable for
any $y$. The effective first-order correction to the Gaussian-like
profile is analytical bounded, which can be obtained by cutting the
infinity from the numerical solution of Eq. (23). These will be
illustrated numerically as follows.

As an instance, setting the parameters
$V_0=V_{01}=3/2,V_1=0.05,k=1.5,g_{1d}=-1,\mu_0=-2$ and the boundary
condition which is equivalent to $A=-\infty, B=0$, from Eqs. (11)
and (23) we plot the chaotic atomic density
$|\psi|^2=(\psi_0+\psi^{e}_{11})^2$ as in Fig. 5a. Here the solid
and dashed curves correspond to $x_c=1, \mu_1=-0.02148$ and $x_c=2,
\mu_1=0.02083$ respectively, which satisfy the generalized Melnikov
chaos criterion (18) and (24) approximately. The dashed curve has
approximate shape with the solid one and can be regarded as the
latter after a translation of distance $1$. The numerically
unbounded first corrections are uncomputable for sufficiently large
$|y|=|x-x_c|$ values and the starting points of the incomputability
are shown to be about $y=\pm y_s\approx \pm 2$ after which the
atomic densities may be irregular and tend to infinity rapidly. By
using the wide-black curves instead of the infinity in range
$|x-x_c|\ge 2$ of Fig. 5a, we obtain the Gaussian-like wave packets
as in Fig. 5b which describe the analytically bounded atomic density
better. The wide-black parts are the sketch maps of the chaotic
regions of density distributions, whose width varies from the
maximal value $\delta(y_s)\sim V_1$ to minimal one $\delta(\pm
\infty)=0$. In the chaotic regions of density, the atomic density is
unpredictable. The effective first corrections in the range
$x\in(-0.5,3)$ are exhibited by the inset of Fig. 5b, which are
plotted from Eq. (23) for the range $|y|<2$ and the parameters
adopted in Fig. 5a. It should be emphasized that the analytically
bounded chaotic states are discrete and can be manipulated
experimentally by taking the barrier heights $V_{0n}$ in Eq. (17)
discontinuously and adjusting the barrier site $x_c$ continuously.
Particularly, by increasing $x_c$ adiabatically \cite{Pu}, we can
move the Gaussian-like wave packets slowly for the purpose of BEC
transport \cite{Paul}.

\begin{figure}
 \center
\includegraphics[width=1.6in]{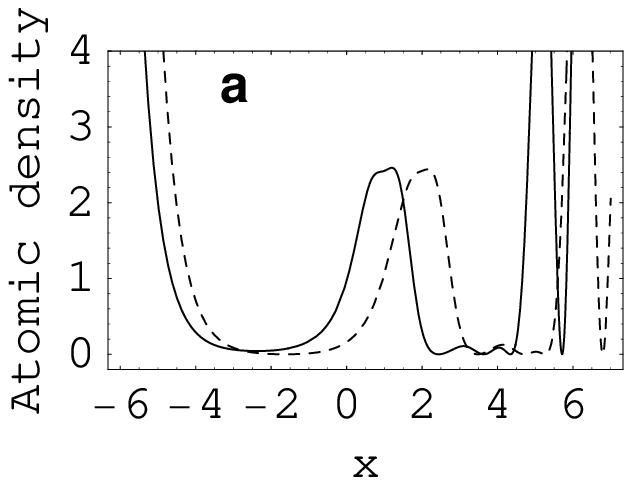}
\includegraphics[width=1.6in]{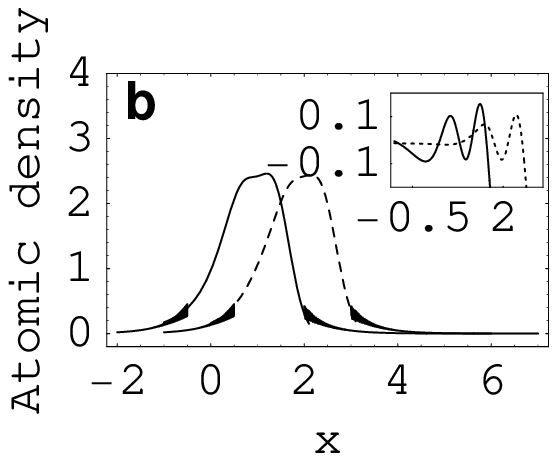}
\caption{(a) The chaotic density profiles of atomic number for
$x_c=1$ (solid curve) and $x_c=2$ (dashed curve). (b) The
analytically bounded density profiles from Fig. 5a by replacing the
parts of $|x-x_c|>2$ with the sketch maps of the chaotic density
regions.}\label{fig5}
\end{figure}

\section{Conclusions and discussions}

We have investigated the spatial structure of the 1D attractive BEC
interacting with a tanh-squared-shaped laser barrier potential and
perturbed by a weak laser standing wave. The existence of the
Smale-horseshoe chaos is demonstrated and the Melnikov chaotic
regions of parameter space are displayed. In the low laser barrier
case, the aperiodic chaotic states and periodic regular states are
illustrated numerically. For the sufficiently strong barrier
potential a set of discrete chaotic solutions is constructed
formally. Any chaotic solution is the combination of a Gaussian-like
wave-packet with the corresponding perturbed correction. The
discrete chaotic solutions are analytically bounded only for the
discrete barrier height values and special magic numbers of
condensed atoms. The density profiles of BEC in the discrete chaotic
states are investigated numerically for the lowest quantum number,
and the numerical instability is revealed. The Gaussian-like wave
could be translated by varying the laser-barrier site adiabatically,
which is similar to the bright soliton of an attractive BEC with the
parabolic barrier potential \cite{Khaykovich}. The periodic
structures of BEC can be detected by the Bragg scattering of an
optical probe beam \cite{Strekalov} and the used Gaussian-like
potential can be generated by a sharply focused laser beam in the
experiments \cite{Burger}. Thus the irregular chaotic states could
be observed and controlled readily with current experimental
capability.

The existence of chaos means the sensitive dependence of the BEC
system on the boundary conditions and parameters in chaotic region.
The sensitivity causes the unpredictability of the spatial
distributions of the BEC atoms, since the boundary conditions cannot
be set accurately in a real experiment. The above results reveal the
possible bounded states associated with the spatial distributions,
and suggest a method to control the irregular chaotic states by
adjusting the lattice strength and laser barrier parameters.

It is worth noting that the discrete chaotic states may appear in
many different physical systems with different Gaussian-like
potentials and may also exist in the temporal and spatiotemporal
evolutions of the time-dependent systems.

\begin{acknowledgments}
This work was supported by the National Natural Science Foundation
of China under Grant Nos. 10575034 and 10875039.
\end{acknowledgments}

\end{document}